# Understanding of double-curvature shaped magnetoimpedance profiles in Joule-annealed and tensioned microwires at 8-12 GHz

V.V. Popov[1], V.N. Berzhansky[1], H.V. Gomonay[2], F.X. Qin[3*]

[1] Taurida National University, Simferopol, Republic Crimea
[2] National Technical University of Ukraine 'KPI', Kyiv, Ukraine
[3] 1D Nanomaterials Group, National Institute for Material Science, Tsukuba, Japan

We have investigated for the first time the combined effect of current and stress on the GMI characteristics of vanishing-magnetostrictive Co-rich microwires at microwave frequency. As the current-annealed wire is subject to certain tensile stress, one can observe a drastic transformation of field dependence of MI profiles from smooth shape of a broad peak to deformed shape of a sharp peak with the emergence of a kink on each side. It follows that three different regions – core, inner and outer shell – have been formed by the combined effect of Joule-annealing, current generated magnetic field and the tensile stress. A critical field sees a drop of field sensitivity from outer to inner shell and shifts to lower value with increasing annealing current. We successfully adapted our core-shell model to a core-shell-shell model by designating different anisotropy field for each region to satisfactorily resolve the unique double-curvature shaped peaks in the field derivative MI profiles.

---

[*] Corresponding author: faxiang.qin@gmail.com

The glass-coated amorphous microwires have drawn much research interest due to its excellent soft magnetic properties, particularly giant magnetoimpedance (GMI) properties and accompanying peculiar magnetic features over a broad frequency range, which allows one probe into the fundamental magnetism and electromagnetics as well as exploit the high-performance sensing applications.[1-3] On the application front, microwires can be conveniently integrated into functional devices at fine scale and give good performance with its excellent magnetic and mechanical properties.[4-6] Co-based glass-coated microwires, among others, draw on its bamboo-like domain structure resulted from the coupling between the negative magnetostriction and frozen-in stress to realize a large and sensitive MI effect for miniature sensors[7,8] It has been well acknowledged that either internal or applied stresses in tensile or torsional mode have a profound effect on the magnetoelastic anisotropy of microwires and hence on GMI.[9,10]

As compared to megahertz frequency, high-frequency GMI behaviour of mirowires remains a great deal to be explored.[9,11-14] In the present work, we have herein directed our efforts into the gigahertz range to emphasize the less explored high frequency physics. Our previous work has proposed a core-shell model to explain the high-frequency magnetic hysteresis induced by the complex stresses.[14] In the current work, we aim to address the co-influence of Joule-annealing and tensile stresses on the magnetoimpedance of Co-based wires via both experimental and modeling approaches. Our results are of vital importance to the design of GMI microwire-based stress sensors.

Two types of Co-based glass-coated amorphous microwires fabricated by a modified Taylor-Ulitovskiy method [15] were studied in the current work: one has a nominal composition $Co_{67.4}Fe_{3.7}Ni_1B_{12}Si_{14}Mo_{1.9}$ with core diameter of 7.2 μm and total diameter of 12μm; the other has a nominal composition of $Co_{66.94}Fe_{3.83}Ni_{1.44}B_{11.51}Si_{14.59}Mo_{1.69}$ with core diameter of 17.4µm and total diameter of 20.4µm. GMI measurements were carried out in the microwave frequency range (8-12GHz) using a waveguide technique, which details has been described in Ref. [16]. The reflection coefficient was measured for the microwire placed in the shorted sample holder. The wire was placed symmetrically in parallel to the narrow wall of the waveguide. The distance from the wire to the metallic short was selected to be about 1cm to ensure the position of the microwire approximately in the electric field antinode at the specific measurement frequency. A current source was connected to the both ends of the tested wire to apply dc currents up to 49mA; the wire is also subject to a constant tensile stress during the measurement. The holder was connected to the conventional scalar network analyzer and we measured the absolute value of the reflection coefficient that is inversely proportional to impedance.[16] Herein the obtained field dependence of reflection coefficient can be dealt with as GMI profiles. External magnetic field was applied along the wire axis using a pair of Helmholtz coils fed from the power amplifier with the quasi-static frequency of 20 Hz. This allows accurate and real-time measurements of the GMI response from the wire.

We found that the application of the current leads to two well-distinguished scenarios depending on the current amplitude. For relatively small currents (0, 4 and 8 mA in Fig. 1a) GMI has a typical 'valve-like' profile corresponding to the longitudinal magnetization process. Additional circular magnetic field of the current enlarges total anisotropy of the sample while preserves smooth shape and amplitude of the peak. Clearly, the small current has typical influence on the GMI features as it was found in earlier works[17]. While the current increases to 12mA, a visible reduction of GMI peak width can be observed, which indicates the initiation of somewhat radical change of the magnetic structure. Indeed, for even larger current (i.e., 15 mA in Fig. 1a) the GMI profile changes abruptly from of broad peak to of sharp peak, suggesting that thermo-magnetic annealing produced by the current significantly modifies the magnetic structure of the wire.

According to the *core-shell model* as we developed in previous work [14], the shape of the GMI peak is mainly determined by the amplitude $H_{sh}$ and direction $\Psi_{sh}$ of the anisotropy field for the shell layer of the wire's metallic core. As-received Co-rich microwire has negative magnetostriction and circular anisotropy in the outer shell, i.e., $\Psi_{sh} = \pi/2$, which results in a smooth shape of the GMI profile. The circular magnetic field generated by the weak current $H_{\varphi}$ contributes additively to the total energy, but has little influence on $\Psi_{sh}$ and hence the broad peak feature remains. To understand sharp peaks observed for large currents, we propose that they are attributed to the compositional and topological short range ordering processes activated during Joule-heating and the magnetostriction constant of the wire increasing towards positive values.[18,19] As thus, a transformation of circular anisotropy $\Psi_{sh} = \pi/2$ to axial anisotropy $\Psi_{sh} = 0$ takes place with the cooperation of current and stress applied.

Figure 1b shows the 'residual' current effect on GMI, which was obtained after we temporarily switched off the current source. As with the GMI profiles before cutting off the current source, application of current till 8mA has invisible influence. This clearly indicates that the concerned current effect is reversible. In contrast, a close inspection of the profiles over 12mA reveals an irreversible distortion that evolves with increasing current. Overall, the broad-to-sharp change transformation and corresponding threshold current value remains the same. Together, it can be inferred that the relatively large current has not changed the basic magnetic configuration consisting of core and shell, but have induced some significant local changes that are reflected in the distortions in the GMI curves.

To further investigate such extraordinary behaviour, we turn to a thicker wire that is able to take much larger current in order that more significant changes can be detected; the results are summarized in Fig.2. As expected, two significant kinks appear on both shoulders of the GMI curves with the annealing current from 13mA to 49mA, which is shaped as a double-curvature in the derivative GMI profiles and its current dependency are better seen as shown in Fig.2b. By extracting the two critical magnetic field values corresponding to the two peaks in the derivative profiles, one can sees from Fig.2c a drop of them with the increasing current. To understand these interesting results, we attempt to find their origin from the modification of the core-shell structure. Since there emerge additional kinks that discontinue the GMI curves, we assume that there should appear in the wire an additional region with its own anisotropy field coupled to the existing core and shell regions. A relatively large dc current can generate strong circumferential anisotropy, which is further reinforced by the applied tensile stress. Joule heat produced by current permeates through the entire wire volume, while only the wire surface is practically cooled down to the ambient temperature during the GMI measurements. The outcome of such combined effect is then split for different regions along the radial direction and the outer shell is accordingly sub-divided into two sections. Thus, three distinct regions are formed in the annealed wire, namely, core with axial anisotropy, inner shell and outer shell both with circumferential anisotropy but having significantly different absolute values of the effective anisotropy fields.

To validate such assumption of 'core- inner shell-outer shell-domain', we introduced additional effective anisotropy to the 'core-shell' model. As a result, the free energy for a single microwire is expressed as

$$F = \kappa_1 \cdot \left(\frac{1}{2}H_1 \sin^2(\psi_1 - \theta_1) - H_z \cos\theta_1\right) + \kappa_2 \cdot \left(\frac{1}{2}H_2 \sin^2(\psi_2 - \theta_2) - H_z \cos\theta_2\right)$$
$$+ \kappa_c \left(\frac{1}{2}H_c \sin^2(\psi_c - \theta_c) - H_z \cos\theta_c\right)$$
$$- H_{int}\cos(\theta_1 - \theta_c) - H_{int}\cos(\theta_2 - \theta_c) - H_{int}\cos(\theta_1 - \theta_2) \quad (1)$$

where $H_1$, $H_2$ are the effective anisotropy field of inner and outer shell, respectively; $\Psi_1$, $\Psi_2$ are the direction of the anisotropy axis in the inner shell and outer shell ranging from 45° to 90°. $H_c$ is the anisotropy field of the core, $\Psi_c$ is the direction of the easy axis in the core that is close to 0°, $H_z$ is the external magnetic field, $\kappa_1$, $\kappa_2$, $\kappa_c$ are respectively the relative volume of the inner shell, outer shell and core with respect to the volume of the wire ($0<\kappa<1$), $H_{int}$ is an effective interaction between the regions. It should be remarked that the equilibrium values of $\theta_c$ and $\theta_{1(2)}$ are different, in that during annealing non-homogeneous field and temperature distribution within the wire sample gives rise to complex distribution of anisotropy field accordingly. According to Ref. [20], GMI response in microwave frequency band is mainly determined by the equilibrium direction of the magnetization at the wire surface and can be computed as $GMI \propto \cos(\theta_2)$. Herein, by minimizing Eq.(1) with respect to angles $\theta_c$ and $\theta_{1(2)}$ (the direction of the magnetization in the core and the inner (outer) shell, one can theoretically model the GMI response of the wire for each value of the external field.

By assigning physically reasonable values of the parameters into the model, one has received good agreement between our modeling and experimental results and produced the major feature of the double-curvature shape, as shown in Fig.3. The validity of the proposed model can also be evidence by the relation between the lower critical field $H_1$ and current as shown in Fig.2c. According to the model, $H_1$ should be the minimum value conditioned by the skin effect whereby the applied magnetic field can reach the boundary between outer and inner shell. As the outer shell volume will be reduced with increasing current, $H_1$ will then decrease. This reasoning coincides well with the experimental results. Therefore, the updated 'core-shell-shell' model successfully addressed the combined effects of current and stress on the microwave GMI behavior of Co-based microwires. On one hand, we can further exploit the model to understand various behavior of microwave GMI under complex fields. On the other hand, the magnetic anisotropy or domain structure can be tailored in a more controllable fashion with our continuously improved understanding of any peculiar features occurred in the high-frequency GMI profiles.

**Acknowledgements**

FXQ is supported by JSPS fellowship and Grants-in-Aid for Scientific Research No. 25-03205.

**Figure captions**

FIG. 1 (a) Effect of dc current from 0 to 15mA on GMI curves for $Co_{67.4}Fe_{3.7}Ni_1B_{12}Si_{14}Mo_{1.9}$ microwire in the presence of a tensile stress of 170MPa; (b) Effect of annealing current from 0 to 17mA on GMI curves for the same microwire in the presence of a tensile stress of 170MPa.

FIG. 2 (a) Experimental field dependence of GMI for $Co_{66.94}Fe_{3.83}Ni_{1.44}B_{11.51}Si_{14.59}Mo_{1.69}$ microwire annealed from 0 to 49 mA in the presence of 300MPa; (b) field dependence of derivative GMI profiles; (c) current dependence of critical field $H_1$ and $H_2$ that are extracted from (b).

FIG. 3 (a) Modelled GMI profiles for the $Co_{66.94}Fe_{3.83}Ni_{1.44}B_{11.51}Si_{14.59}Mo_{1.69}$ microwire annealed by 37 mA in the presence of 300MPa. The parameters used for simulation are as follows: $H_1 = 0.02$ Oe, $H_2 = 3.0$ Oe, $H_c = 0.4$, Oe, $\Psi_c = 0°$, $\Psi_{1,2} = 90°$, $H_{int} = 0.5$ Oe, $\kappa_c = 0.8$, $\kappa_c = 0.15$, $\kappa_c = 0.05$. The inset shows the derivative GMI profiles (b) the corresponding experimental results singled out for comparision.

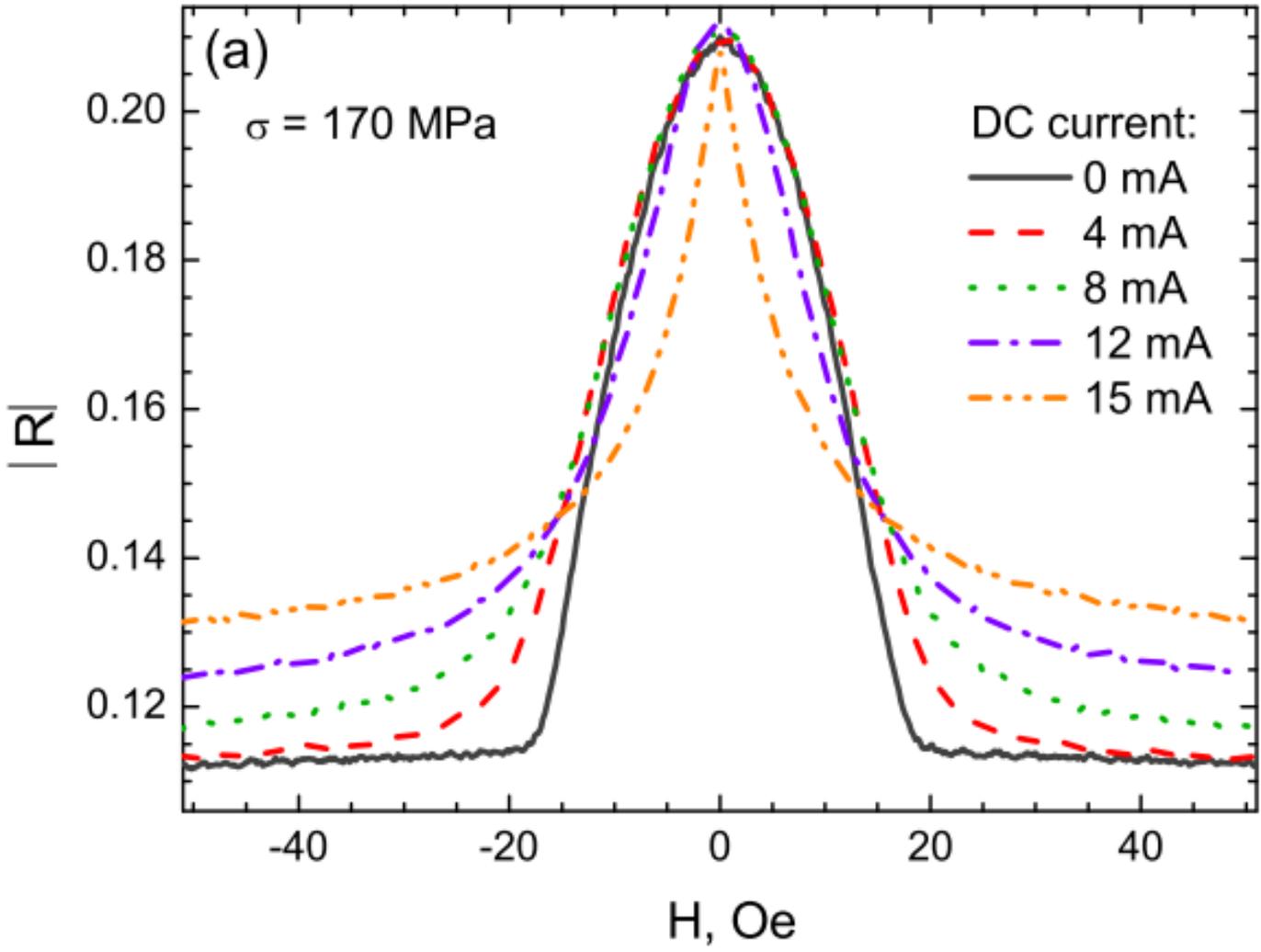

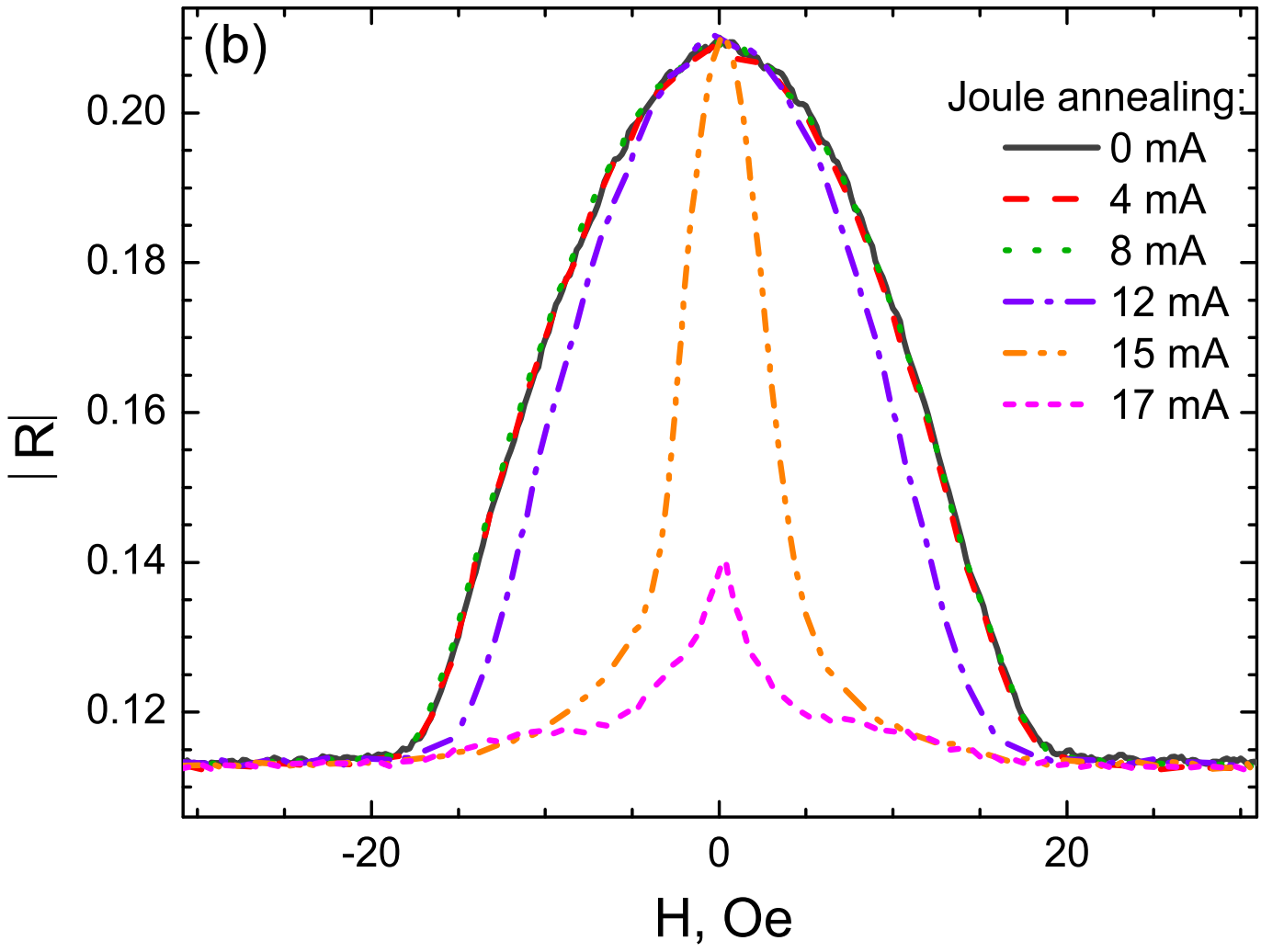

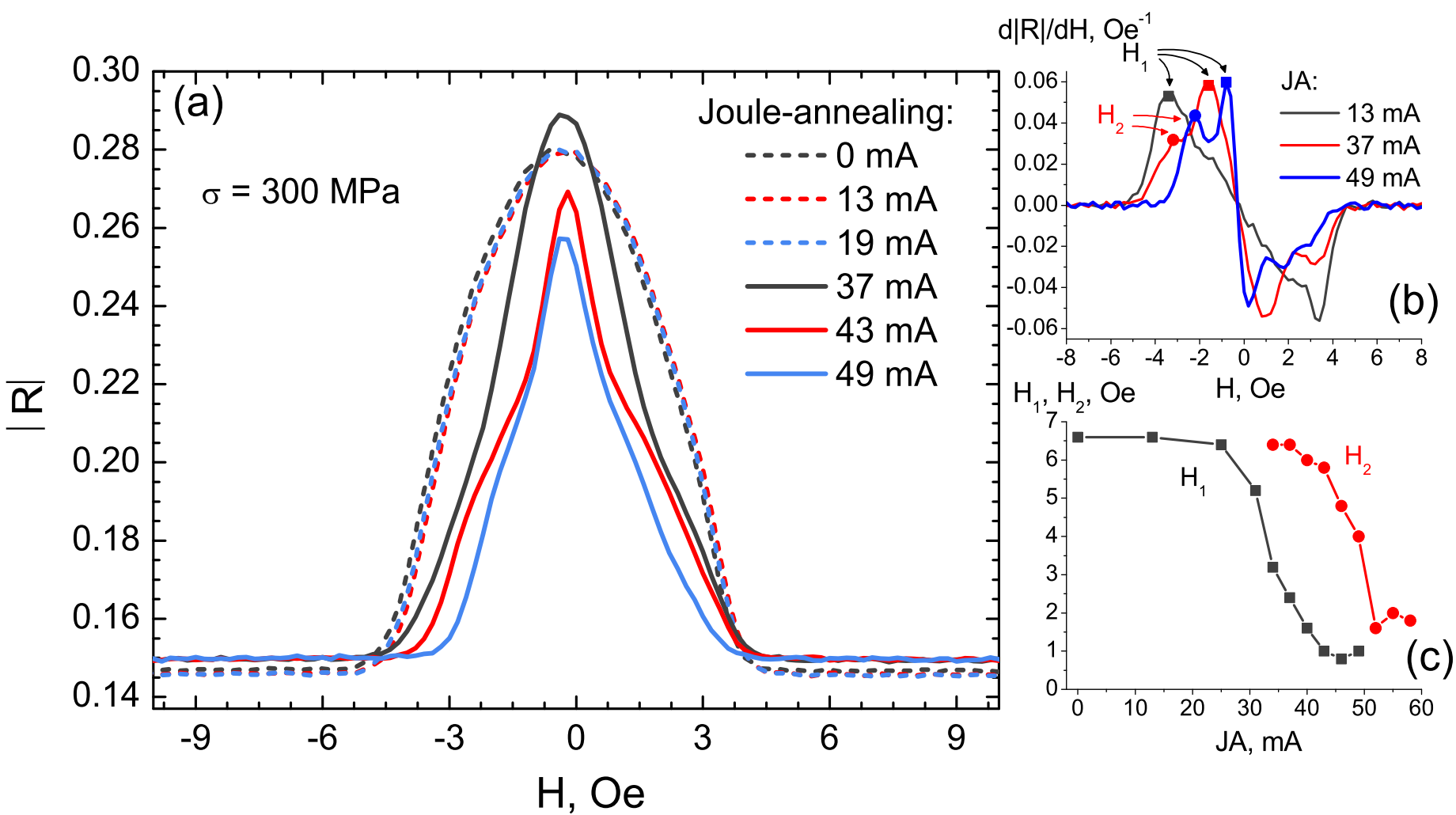

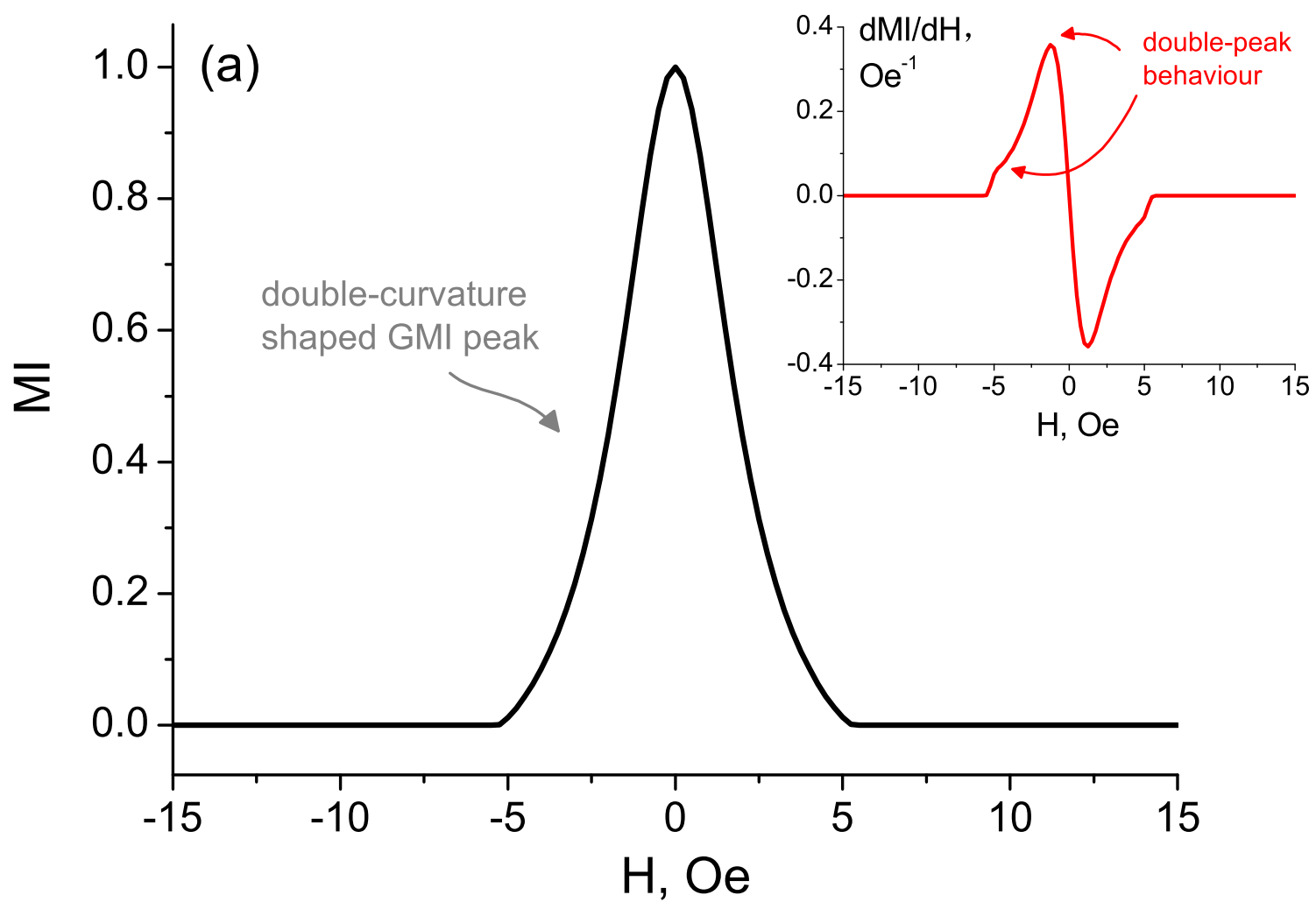

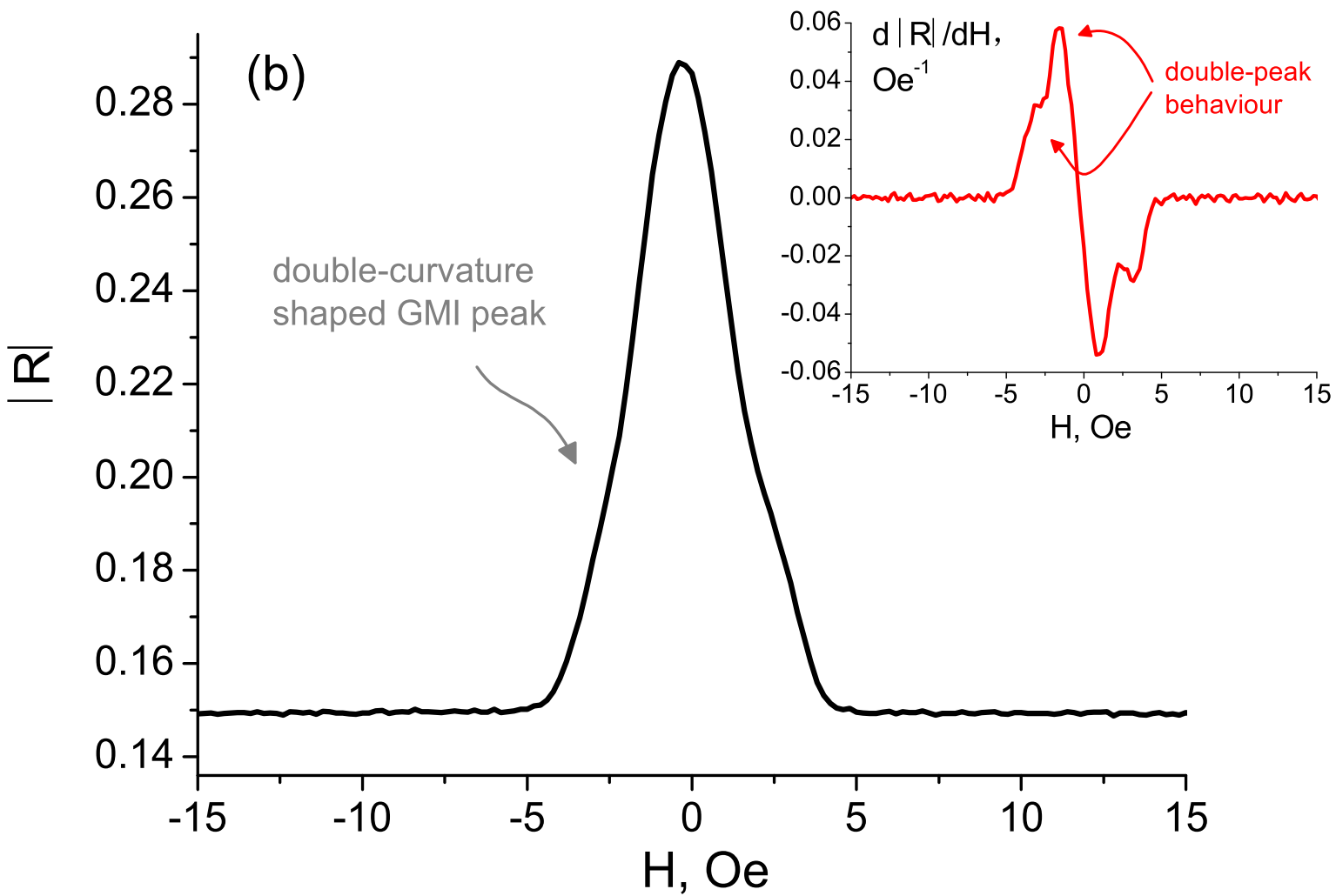